\documentclass[12pt]{elsart}

\usepackage{graphicx,epsfig,amsmath,amssymb,amsfonts}
\usepackage{latexsym,cite,eucal}
\newcommand{\beq}{\begin{equation}}
\newcommand{\eeq}{\end{equation}}
\newcommand{\nn}{\nonumber \\}
\newcommand{\fs}{\,.}
\newcommand{\co}{\,,}

\renewcommand{\dag}{^\dagger}

\newcommand{\eps}{\epsilon}

\newcommand{\mr}{\mathrm}

\newcommand{\MeV}{\,\mbox{MeV}}

\newcommand{\mpn}{M_{\pi^0}}
\newcommand{\mpc}{M_\pi}
\newcommand{\pin}{\pi^0}

\renewcommand{\vec}[1]{{\bf #1}}

\newcommand{\po}{p_2}
\newcommand{\pn}{p_1}

\newcommand{\Fnr}{\mathcal{F}}
\newcommand{\mpr}{m_p}
\newcommand{\mn}{m_n}
\renewcommand{\O}{O}
\newcommand{\vq}{{\bf q}}
\newcommand{\vk}{{\bf k}}
\newcommand{\kt}{k_0}
\newcommand{\knl}{k_1}
\newcommand{\qstar}[1]{h^{#1}(s,m_c,M_{\pi^d})}

\begin{document}

\begin{frontmatter}

\title{\Large\bf Pion photoproduction in a nonrelativistic theory}

\author{Andreas Fuhrer}

\address{University of California, San Diego
\\9500 Gilman Drive, La Jolla, CA 92093--0319}

\thanks[fuhrer]{afuhrer@physics.ucsd.edu}

\begin{abstract}
The pion and nucleon mass differences generate a very
    pronounced cusp in the photoproduction reaction of a single pion on the
    nucleon. A nonrelativistic effective field theory to
    describe this reaction is constructed. The approach is rigorous in the sense that it is an effective
field theory with a consistent power counting scheme. 
Expressions for the $S$- and $P$-wave multipole amplitudes for all
four reaction channels at two loops are given.
\end{abstract}

\end{frontmatter}

{\bf 1.} Isospin breaking effects have recently received a considerable amount
of attention. This was triggered by the observation of a cusp-like structure in
the decay rate of $K \to 3\pi$ decays \cite{NA48cusp}, which
originates from isospin breaking. Since the strength of this cusp is
intimately related to the $\pi\pi$ scattering lengths, another
possibility to measure the $\pi\pi$-scattering lengths -- beside
$K_{e4}$ decays \cite{E865,NA48ke4}  and the lifetime of the pionium atom \cite{dirac} -- was
established. It is amusing to note that the same effect that is
responsible for the cusp in $K \to 3\pi$ decays also leads to
a correction to the $\pi\pi$-scattering phase shifts from
$K_{e4}$ decays \cite{CGR}.\newline
The photoproduction reaction of neutral pions on the proton is a
reaction where isospin breaking corrections have not yet been
calculated in a fully systematic approach. On the other hand, it
has been known for a long time that this reaction shows a very strong effect
due to isospin breaking: the electric multipole $E_{0+}$ exhibits an
exceptionally strong cusp at the $\pi^+ n$ threshold (see for instance
Ref.~\cite{bw}). In much the same
way as in $K \to 3\pi$ decays, a nonrelativistic theory -- adapted in
this letter to the photoproduction reaction of pions on nucleons -- provides a rigorous
framework which yields quantum field theoretical matrix elements
with the effective range parameters of pion-nucleon scattering and the threshold
parameters of the photoproduction reaction as free
parameters. Furthermore, the framework allows one to include radiative corrections in a
standard manner at a later stage.

The cusp in neutral pion photoproduction has been studied
before. Ref.~\cite{BKM2} introduces a two-parameter model, which captures
the most important leading effect of the cusp. In
Ref.~\cite{bernstein}, a coupled channel $S$-matrix approach is used
to investigate the cusp structure.

In a first step, the nonrelativistic Lagrangian is written down and the power counting
is discussed. Then, the coupling constants are matched to the pertinent threshold
parameters. The calculation of the multipole amplitudes up to two-loop
order is then straightforward. The representation of the amplitudes
correctly reproduces the analytical structure in the low energy
region. As an application, the relation of the phase of the electric
multipole $E_{0+}$ to the phase of the $S$-wave of $\pi^0 p \to \pi^0
p$ scattering is discussed. 

{\bf 2.} Some basic relations and definitions used in the
analysis of pion photoproduction are collected.
We calculate the matrix element for the process  $N(p_1)+\gamma(k) \to
N(p_2) + \pi^{0,\pm}(q)$ for all four channels at leading order in the
electromagnetic coupling $e$, 
\begin{align}
\langle p_2, q\,\, \mathrm{out} | p_1, k\,\, \mathrm{in} \rangle &=  -i (2\pi)^4
\delta^{(4)}(P_f-P_i) \, \bar{u}(\po,t') \eps_\mu J^\mu u(\pn,t) \co
\end{align}
where $N$ denotes either a proton or a neutron, $P_i$ and $P_f$ denote
the total four momentum in the initial 
and in the final state, respectively and $\eps^\mu$ stands for the
polarization vector of the photon. In the following, the four reaction
channels will be abbreviated as
\begin{align}
\gamma p \to p \pi^0 &: (p0) \co &\gamma p \to n \pi^+ &: (n+) \co
&\gamma n \to n \pi^0 &: (n0) \co &\gamma n \to p \pi^-  &: (p-)\fs \nonumber
\end{align}
To analyze the photoproduction reaction of pions, electric
and magnetic multipoles are usually introduced. To this end, the amplitude
is written in terms of two component spinors $\xi_t$ and Pauli matrices
$\tau^k$ \cite{CGLN},
\begin{align}\label{eq:nrr}
\CMcal{M} &= 8\pi \sqrt{s}\, \xi\dag_{t'}\, \Fnr\,
\xi_{t} \co \nn
\Fnr &= i \boldsymbol\tau \cdot \boldsymbol\epsilon\,  \Fnr_1+
\boldsymbol\tau\cdot \hat{{\bf q}}\, \boldsymbol\tau\cdot (\hat{{\bf k}}\times
\boldsymbol\epsilon)\,\Fnr_2 +i \boldsymbol\tau\cdot \hat{{\bf k}}\,
\hat{{\bf q}}\cdot \boldsymbol\epsilon\, \Fnr_3 +i \boldsymbol\tau
\cdot\hat{{\bf q}}\, \hat{{\bf q}}\cdot \boldsymbol\epsilon\, \Fnr_4
\fs
\end{align}
The hat denotes unit vectors. The $\Fnr_i$ are decomposed into
electric and magnetic multipoles with the help of 
derivatives of the Legendre polynomials $P_l(z)$ \cite{CGLN},\newpage
\begin{align}
\Fnr_1 &= \sum_{l=0}\,
    [lM_{l+}+E_{l+}]P_{l+1}'(z)+[(l+1)M_{l-}+E_{l-}]P_{l-1}'(z) \co \nn
\Fnr_2 &= \sum_{l=1}\,[(l+1)M_{l+}+lM_{l-}]P_l'(z) \co \nn
\Fnr_3 &=
\sum_{l=1}\,[E_{l+}-M_{l+}]P_{l+1}''(z)+[E_{l-}+M_{l-}]P_{l-1}''(z)
\co \nn
\Fnr_4 &= \sum_{l=1}\,[M_{l+}-E_{l+}-M_{l-}-E_{l-}]P_l''(z) \fs
\end{align}
The discussion is restrained to the
center of mass frame in the rest of the article.

{\bf 3.} To describe the behavior of the multipoles close to threshold -- where the energy of the produced
pion and of the nucleon are small -- a nonrelativistic
calculation is justified. Furthermore, it offers the advantage that
all the masses can be set to their physical value. Therefore, all the
poles and branch points appear at the correct place in the Mandelstam
plane. Moreover, the interaction of the nucleon and the pion is described by effective
range parameters, which allows one to directly access the pion-nucleon scattering
lengths.

The covariant formulation of nonrelativistic field theories introduced in\linebreak
Refs.~\cite{CGKR,BFGKR,KS} is used here since it incorporates the correct
relativistic dispersion law for the particles.  The nonrelativistic proton, neutron and pion fields are denoted by
$\psi$, $\chi$ and $\pi_k$, respectively. The kinetic part of the Lagrangian after minimal substitution takes
the form (see Ref.~\cite{BFGKR2})
\begin{align}
\CMcal{L}_{kin} &= \sum_\pm\Bigl(i\pi_\pm^\dagger D_t{\cal W}_\pm\pi_\pm
-i(D_t{\cal W}_\pm\pi_\pm)^\dagger\pi_\pm-2\pi_\pm^\dagger {\cal W}_\pm^2\pi_\pm\Bigr)\nn
&+ i\psi^\dagger D_t{\cal W}_p\psi
-i(D_t{\cal W}_p\psi)^\dagger\psi-2\psi^\dagger {\cal W}_p^2\psi \nn
&+ 2 \chi\dag W_n (i \partial_t-W_n)\chi + 2 \pi_0\dag W_0 (i\partial_t-W_0)\pi_0 \co
\end{align}
with
\begin{align}
W_0 &= \sqrt{\mpn^2-\triangle} \co &W_n &= \sqrt{m_n^2-\triangle}\co
&D_t\pi_\pm &= (\partial_t\mp ieA_0)\pi_\pm \co\nn
D_t\psi &= (\partial_t- ieA_0)\psi \co &{\cal W}_\pm &= \sqrt{M_\pi^2-{\bf D}^2} \co &{\cal W}_p &=
\sqrt{\mpr^2-{\bf D}^2} \co \nn 
{\bf D}\pi_\pm &= (\nabla\pm ie{\bf A})\pi_\pm \co &
{\bf D}\psi &= (\nabla + ie{\bf A})\psi \fs
\end{align}
Note that since the photon is treated as an external field, its
kinetic term is absent.

{\bf 4.} Consider the reaction\footnote{The arguments remain of course the same for the
  other three channel.} $p(p_1)+\gamma(k) \to p(p_2) + \pin(q)$. Close
to threshold, the momenta of the incoming proton and photon are of the 
order of the pion mass whereas the outgoing particles have very
small momenta. Therefore, we count momenta of the outgoing pion and
the outgoing proton as a small quantity of
$\O(\eps)$ and the momenta of the incoming proton and of the photon as
$\O(1)$. All the masses are counted as $\O(1)$. The mass differences
of the charged and neutral pion, $\Delta_\pi \equiv \mpc^2-\mpn^2$ and of
the proton and the neutron, $\Delta_N \equiv \mn^2-\mpr^2$ are counted as $\O(\eps^2)$. We refrain from counting
the pion mass as a small quantity with respect to the nucleon mass,
since considering a particle as light and nonetheless describing it
nonrelativistically seems inconsistent  (see also Ref.~\cite{zemp}).\newline
At first sight, this counting scheme seems to lead to infinitely many terms already in the leading
order $p +\gamma \to p +\pin$ Lagrangian $\CMcal{L}_\gamma$ because derivatives on the
incoming fields are not suppressed in $\eps$.
However, since the the modulus of the momentum of the incoming
particles, $|{\bf k}|$, can be expanded in the small momentum $|{\bf q}|$,
\begin{align} \label{eq:k}
|{\bf k}| &= \sum_{n}k_n \vq^{2n} \co &\kt &=
\frac{\mpn}{2}\, \frac{2+y}{1+y} \co &\knl &= \frac{y^2+2y+2}{4 \mpn (1+y)} \co &y &=
\frac{\mpn}{\mpr} \co
\end{align}
one obtains a valid power counting scheme: Consider the Feynman
rule in momentum space of an operator of order
$\O(\eps^0)$  with a given arbitrary number of derivatives. This
expression can be expanded
in a sum of {\it one} term of order $\O(\eps^0)$ without any momenta
of the incoming particles present and subsequent higher order
terms. Doing this for every operator of $\O(\eps^0)$, all the resulting
leading order terms without any momentum dependence can be described
by one operator of order $\O(\eps^0)$ in the interaction Lagrangian. The
same procedure leads to finite numbers of operators at any given higher order
in $\eps$. The derivatives on the incoming fields are only
needed to generate unit vectors in the direction of the incoming
photon. This shows that the nonrelativistic theory is not capable of predicting
the dependence on $|{\bf k}|$ even at threshold. \newline
An additional generic parameter $a$ is introduced to count the
pion-nucleon scattering vertices. Every pion-nucleon interaction
vertex counts as a quantity
of order $\O(a)$ since the coupling constants are proportional to the
pion-nucleon scattering threshold parameters, which are
small.
The perturbative expansion is therefore a combined
expansion in $\eps$ and $a$.

{\bf 5.} The Lagrangian needed for the calculation of the amplitudes for pion 
photoproduction reads $\CMcal{L} = \CMcal{L}_{kin}+\CMcal{L}_\gamma+\CMcal{L}_{\pi N}$,
where $\CMcal{L}_{kin}$ denotes the kinetic part, $\CMcal{L}_\gamma$
incorporates the interaction with the photon field and and
$\CMcal{L}_{\pi N}$ describes the
pion-nucleon sector.

In the pion nucleon sector, the leading terms of the Lagrangian have
been given before in Ref.~\cite{LR}. First, some notation is introduced in order to
write the Lagrangian in a compact form. 
For every channel $n$, we collect the charges of the outgoing and the
incoming pions in the variables $v$ and $w$, $(n;v,w)$: $(0;0,0) ,
(1;0,+),\, (2;+,+),\, (3;0,0),\, (4;-,0),\, (5;-,-)$, thereby assigning \linebreak
unique values to the variables $v$ and $w$ once $n$ is given. The
Lagrangian reads
\begin{align}\label{eq:Lpn}
\CMcal{L}_{\pi N} &= \left( \psi\dag \,\, \chi\dag \right)
\left( \begin{array}{cc} T_{\{0,5 \}} & T_{\{1,4\}}
  \\ T_{\{1,4\}}\dag & T_{\{2,3 \}} \end{array}
\right) \left( \begin{array}{c} \psi \\ \chi \end{array} \right) \co
\\[2mm]
T_\CMcal{C} &= \sum_{n \in\, \CMcal{C}}\left[ C_n \pi_v\dag \pi_w + D_n^{(1)}\nabla^k
\pi\dag_v \nabla^k \pi_w + D_n^{(2)} \pi\dag_v
\overleftrightarrow{\triangle} \pi_w + i D_n^{(3)} \tau^k
\epsilon^{ijk}\nabla^i \pi\dag_v \nabla^j \pi_w \right]  \nonumber
\end{align}
with the abbreviation $f\overleftrightarrow{\triangle}g \equiv f \triangle
g + (\triangle f) g$.\newline
For $\CMcal{L}_\gamma$, the photon is treated as an external vector field ${\bf A}$ which is
odd under parity and time-reversal transformations. Gauge invariance
requires that it can 
only appear in covariant derivatives and through the Maxwell equations in
the electric and magnetic field ${\bf E} =
-\boldsymbol\nabla A^0 - \dot{{\bf A}}$ and ${\bf B} =
\boldsymbol\nabla \times {\bf A}$. 
The interaction Lagrangian can be read off from the nonrelativistically
reduced and multipole expanded matrix element, Eq.~(\ref{eq:nrr}) and 
the known threshold behavior of the multipoles, $E_{l\pm},M_{l\pm} \sim |\vq|^l$. 
As already mentioned, the effective theory fails to predict the
dependence on $|\vk|$ even at threshold. Factors of $|\vk|$ can always
be obtained by a redefinition of the coupling constants of the Lagrangian. 
Of course, all the terms have to be invariant under space rotations,
parity and time reversal transformations. 
This leads to the following expression for the Lagrangian
$\CMcal{L}_\gamma$. The upper index on the coupling constants counts
the number of derivatives on the external vector field and is introduced for later convenience.
\begin{align}
\CMcal{L}^{(0)}_\gamma &= -i G_0^{(1)}\psi\dag \tau^k \psi\, E^k\,
\pi_0\dag \nn
\CMcal{L}^{(1)}_\gamma &= -i G_1^{(2)}\, \psi\dag \tau^k
\psi\, \nabla^j E^k\, \nabla^j \pi_0\dag  + i G_2^{(1)}\, \psi\dag \tau^m \tau^l
\psi\,B^l\, \nabla^m \pi_0\dag \nn
 &-i G_3^{(2)}\, \psi\dag \tau^j \psi\,
\nabla^j E^k\, \nabla^k \pi_0\dag  \nn
\CMcal{L}^{(2)}_\gamma &= -iG_4^{(3)} \psi\dag \tau^k \psi \nabla^{jl} E^k
\nabla^{jl} \pi_0\dag -i G_5^{(1)} \psi\dag \tau^k \psi E^k \triangle \pi_0\dag \nn
&+i G_6^{(2)} \psi\dag \tau^m \tau^l \psi \nabla^{n} B^l \nabla^{mn}
\pi_0\dag  -i G_7^{(3)} \psi\dag \tau^j \psi \nabla^{jl} E^k \nabla^{kl}
\pi_0\dag \nn 
&-i G_8^{(1)} \psi\dag \tau^j \psi E^k \nabla^{jk}\pi_0\dag \nn
\CMcal{L}^{(3)}_\gamma &= -iG_9^{(2)} \psi\dag \tau^k \psi \nabla^j E^k
\triangle \nabla^j \pi_0\dag -i G_{10}^{(4)} \psi\dag \tau^k \psi
\nabla^{lmn}E^k \nabla^{lmn} \pi_0\dag \nn &+ i G_{11}^{(1)} \psi\dag \tau^m
\tau^l \psi B^l \triangle \nabla^m \pi_0\dag + i G_{12}^{(3)}
\psi\dag \tau^m \tau^l \psi \nabla^{in} B^l \nabla^{min} \pi_0\dag
\nn
&-i G_{13}^{(2)} \psi\dag \tau^j \psi \nabla^j E^k \triangle \nabla^k
\pi_0\dag -i G_{14}^{(4)} \psi\dag \tau^j \psi \nabla^{jlm} E^k \nabla^{klm}
\pi_0\dag \nn
&-i G_{15}^{(2)} \psi\dag \tau^j \psi \nabla^l E^k \nabla^{jkl}
\pi_0\dag 
\end{align}
Here, the notation $\nabla^{i_1i_2\ldots i_k} \equiv \nabla^{i_1}
\nabla^{i_2} \cdots \nabla^{i_k}$ is used. Since the structure of the
Lagrangian for the remaining channels stays the same, one only has to
replace the coupling constants and the field operators,
\begin{align}\label{eq:rep}
(n+) &: \{\psi\dag,\pi_0\dag,G_i^{(n)}\} \to \{\chi\dag,\pi_+\dag,H_i^{(n)} \} \co
  &(n0) &: \{\psi,\psi\dag, G_i^{(n)} \} \to \{ \chi,\chi\dag,L_i^{(n)}\}  \co \nn
  (p-) &: \{\psi,\pi_0\dag,G_i^{(n)} \}   \to \{\chi,\pi_-\dag,  K_i^{(n)}\}
  \fs
\end{align}
The full interaction Lagrangian $\CMcal{L}_\gamma$ is then given by
adding the $\CMcal{L}^{(i)}_\gamma$ of all four channels.

Here, a remark about the structure of the interaction terms seems in
order. Since the interaction Lagrangian contains by construction no $\pi
NN$ couplings, there is no nucleon pole diagram.
\begin{figure}
\centering
\includegraphics[width=9cm]{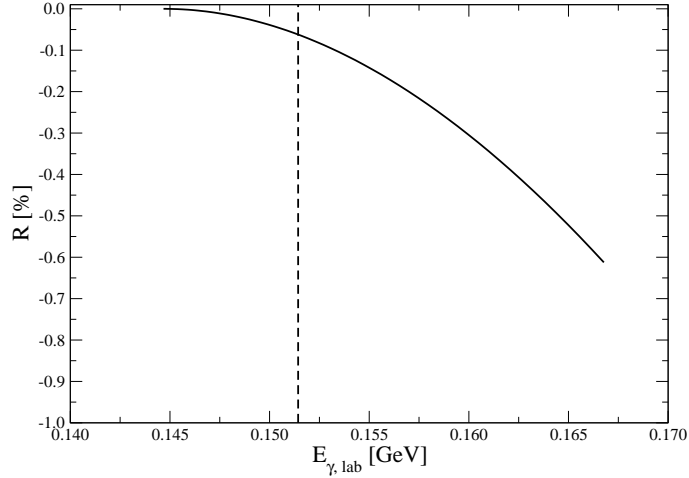}
\caption{The ratio $R$ plotted as a function of the photon energy in
  the lab frame. The dashed line shows the $n \pi^+$ threshold.}\label{fig:approx}
\end{figure}
Therefore one might be worried that an expansion in powers of the momentum $\vq$
does not converge sufficiently fast even in a vicinity of the
threshold. However, we check explicitly that the nucleon pole in the
$s$ channel can be approximated by a polynomial in
$\vq$. 
The tree level expressions of $E_{0+}$ in the
channel $(p0)$ (taken from Ref.~\cite{BKM2}) expanded up to and including terms of
order $\vq^2$ yields a good approximation of the full tree level
result. In Fig.~\ref{fig:approx}, the ratio $R =
(E_{0+}^{\mathrm{exp.}}-E_{0+})/E_{0+}$ is plotted as a function of
the photon energy in the rest frame of the proton.
In view of this result, we also conclude that there
is no reason to be concerned about singularities from resonances
like the $\rho$ in the $t$ channel or the $\Delta(1232)$ in the $s$
channel, which are almost as close to threshold.

{\bf 6.} In the pion-nucleon sector, the coupling constants of the
nonrelativistic Lagrangian, $C_i$ and $D^{(k)}_i$ can be expressed in
terms of pion-nucleon scattering lengths of the $S$-wave and $P$-wave,
$a_{0+}$ and $a_{1\pm}$ and effective range parameters $b_{0+}$, respectively. Adopting the notation of
Ref.~\cite{Hohler},
in the isospin limit, the isospin decomposition of the $\pi N$ scattering amplitudes reads
\begin{align}\label{eq:amps}
T_{p\pi^0\to p\pi^0} &= T_{n\pi^0 \to n\pi^0} = T^+ \co &T_{p \pi^0
  \to n\pi^+} &= T_{n\pi^0 \to p
  \pi^-} = -\sqrt{2}\, T^-
\co \nn T_{n\pi^+ \to n \pi^+} &= T_{p\pi^- \to p\pi^-} = T^++T^- \fs
\end{align}
Defining $\CMcal{N} = 4\pi (\mpr+\mpc)$, one finds
\begin{align}\label{eq:matching}
C_0 &= 2\, \CMcal{N}  a_{0+}^+ \co &C_1 &= 2\sqrt{2}\,\CMcal{N}
a_{0+}^-  \co &C_2 &= 2\, \CMcal{N} (a_{0+}^++a_{0+}^-) \co\nn
C_3 &= C_0 \co &C_4 &= C_1 \co &C_5 &= C_2 \fs
\end{align} 
The matching conditions for the $D_i^{(k)}$ are given in a generic
form only. The isospin index of the threshold parameters can be
inferred from Eq.~(\ref{eq:amps})\footnote{Note that we use the Condon-Shortley phase
convention.}.
\begin{align}
D^{(1)}_i &= 2\CMcal{N} (2 a_{1+}+a_{1-}) \co &D^{(2)}_i &= -
\CMcal{N} \left(\frac{a_{0+}}{2\mpr \mpc}+b_{0+} \right) \co \nn
D^{(3)}_i &= 2\CMcal{N}\ ( a_{1-}-a_{1+})\fs
\end{align}
Here, higher order terms in the threshold parameters have been dropped.
The corrections to these relations which appear due to isospin breaking have to be
calculated within the underlying relativistic theory. For the $C_i$, they can be found
in Refs.~\cite{GILMR,MRR,hkm}. Note that the second line in
Eq.~(\ref{eq:matching}) is only true in the isospin limit.

The constants $G^{(n)}_i$, $H^{(n)}_i$, $K^{(n)}_i$ and $L^{(n)}_i$ on the other hand are related to the
threshold parameters of the electric and magnetic multipoles of the
pertinent channel. 
In the isospin limit, the expansion of the real part of the multipole
$X_{l\pm}$ close to threshold is written in the form 
\begin{align}\label{eq:thresholdpar}
\mathrm{Re}X_{l\pm}(s) &= \sum_{k=0}^{\infty} \bar{X}_{l\pm,2k}|\vq|^{l+2k} \co
\end{align}
which defines the threshold parameters $\bar{X}_{l\pm,2k}$. In the
following, the relations of the coupling constants $G^{(n)}_i$ to these threshold
parameters is given at leading order in the pion nucleon threshold parameters. Since the
nonrelativistic theory is not suited for the study of the dependence
of the multipoles on $|\vk|$, in this analysis, all vectors $\vk$ are turned into unit
vectors by the pertinent redefinition of the coupling constants,
\beq\label{eq:redef}
 G_i^{(n)} = \CMcal{N}_0 \kt^{-n} G_i \co \qquad \CMcal{N}_0 = 4\pi (\mpr+\mpn)\fs
\eeq
Note that the higher order corrections due to Eq.~(\ref{eq:k}) are taken care of in the matching relations.
Again, these relations pick up isospin breaking corrections which have to be
evaluated in the underlying relativistic theory.

Only the matching equations for the couplings of the Lagrangians
$\CMcal{L}_\gamma^{(0)}$ and $\CMcal{L}_\gamma^{(1)}$ are indicated
in the main text, the remaining relations are relegated to appendix
\ref{app:matching}. To ease notation, $\bar{X}_{i\pm} \equiv
\bar{X}_{i\pm,0}$ is used.
\begin{align}
G_0 &= 2\bar{E}_{0+} \co  &G_1 &= 6(\bar{E}_{+1}+\bar{M}_{+1})
\co \nn
G_2 &= -2(\bar{M}_{-1}+2 \bar{M}_{+1}) \co  &G_3 &= 6
(\bar{E}_{1+}-\bar{M}_{1+}) \fs
\end{align} 
For the coupling constants $H_i$, $K_i$ and $L_i$, the algebraic form of the relations is identical. However, the multipoles of
the pertinent channels appear and the masses in Eq.~(\ref{eq:redef})
have to be adjusted.

Upon the inclusion of dynamical photons, the nonrelativistic
coupling constants pick up an imaginary part due to $\gamma N$
intermediate states (see Ref.~\cite{zemp} for a discussion of this issue). For the coupling
constants $C_i$, the imaginary part can be obtained from
Ref.~\cite{hkm}. 
The imaginary part of the coupling constants $G_0$ and $H_0$ in the photoproduction Lagrangian
$\CMcal{L}_\gamma$ follows from the imaginary parts of $E_{0+}$ at threshold. A
calculation at {\it leading order} in chiral perturbation theory \cite{GSS}
yields 
\begin{align}
\mathrm{Im}E_{0+} &= \frac{e^3 g\, y(y+2)}{ 64 \pi^2  F
  (1+y)^\frac{11}{2}}
\bigg[1-(y+1)^3\ln(1+y)\bigg]
\end{align}
for the channel  $(p0)$ and 
\begin{align}
\mathrm{Im}E_{0+} &= \frac{e^3 g (y+2)}{\sqrt{2} \, 32 \pi^2  F
  (1+y)^\frac{11}{2}}
\bigg[\frac{1}{12}(2y+5)(y^2+2)-(y+1)^3 \ln(1+y)\bigg]
\end{align}
for the channel $(n+)$.
Here, $F$ and $g$ denote the pion decay constant and the
axial coupling, both in the chiral limit.
A comparison with the real part at threshold -- which is estimated
taking the experimental results from Refs.~\cite{schmidt,korkmaz} -- shows that the imaginary
part is indeed of the generic order of electromagnetic corrections, 
\begin{align}
\frac{\mathrm{Im}\, G_0}{\mathrm{Re}\,G_0} &\simeq O(10^{-2}) \co  &\frac{\mathrm{Im}\,
  H_0}{\mathrm{Re}\,H_0} &\simeq O(10^{-3}) \fs
\end{align}
In the following, we assume that the coupling constants are real.

{\bf 7.} In the following, we provide the expressions for the electric and
magnetic multipoles $E_{l +}$ for $l=0,1$ and $M_{l \pm }$ for $l = 1$
for the channel $(p0)$. 
The result is written in the form 
\begin{align}\label{eq:result}
X_{l,\pm}(s) &= X_{l\pm}^{\mathrm{tree}}(s)+X_{l\pm}^{\mathrm{1
    Loop}}(s) +X_{l\pm}^{\mathrm{2
    Loop}}(s) \cdots
\end{align}
where $s = (p_1+k)^2$ and the ellipsis denote higher order terms in the expansion in
$\eps$ and $a$.
The results for the other channels can be recovered by a simple replacement of the coupling
constants which will be given later. Writing
\begin{align}\label{eq:tree}
X^{\mathrm{tree}}_{l\pm}(s) &= X^t_{l\pm}\vq^l + X^t_{l\pm,2}
\vq^{2+l} + \cdots \fs
\end{align}
one finds
\begin{align}
E^t_{0+} &= G_0 \co & 3 E^t_{0+,2} &=
G_4-3 G_5+G_6-G_8 \co \nn
6 M^t_{1+} &= G_1-G_3 \co &M^t_{1+,2} &=
-\tfrac{1}{6}G_9+\tfrac{1}{10}G_{10}+\tfrac{1}{15}G_{12}+\tfrac{1}{6}G_{13}-\tfrac{1}{30}G_{14} \co
\nn
 3 M^t_{1-} &= G_3-G_1-3G_2 \co & M^t_{1-,2} &=
 \tfrac{1}{3}G_9-\tfrac{1}{5}G_{10}+G_{11}-\tfrac{1}{3}G_{12} \nn
&&& -\tfrac{1}{3}G_{13}+\tfrac{1}{15}G_{14} \co
 \nn
 6 E^t_{1+} &= G_1+G_3 \co &E^t_{1+,2} &=
 -\tfrac{1}{6}G_9+\tfrac{1}{10}G_{10}+\tfrac{1}{15}G_{12}-\tfrac{1}{6}G_{13}\nn
&&& +\tfrac{1}{30}G_{14}-\tfrac{1}{15}G_{15} \fs
\end{align}
One observes that $D$-waves appear naturally at order $\epsilon^2$ in
this framework (see also Ref.\cite{FBD}).

{\bf 8.} All the one-loop contributions are proportional to the basic integral
\begin{align}
J_{ab}(P^2) &= \int \frac{d^Dl}{i(2\pi)^D} \frac{1}{2 \omega_a({\bf
    l})2\omega_b({\bf P}-{\bf l}) }\, \frac{1}{( \omega_a({\bf
    l})-l_0 ) (\omega_b({\bf P}-{\bf l}) -P_0 +l_0 ) } \co \nonumber
\end{align}
\begin{align}
\omega_\pm({\bf  p}) &= \sqrt{\mpc^2+ {\bf p}^2} \co  &\omega_i({\bf p}) &=
\sqrt{m_i^2+ {\bf p}^2}\co \qquad i=n,p  \nn
\omega_0({\bf p})  &= \sqrt{\mpn^2+ {\bf p}^2}\co  &P^2 &= P_0^2-{\bf
  P}^2 \fs
\end{align}
In the limit $D \to 4$,
\beq\label{eq:loopfunc}
J_{ab}(P^2) = \frac{i}{16\pi
  s}\sqrt{(s-(m_a+M_{\pi^b})^2)(s-(m_a-M_{\pi^b})^2)}\co 
\eeq
which is a quantity of order $\eps$.
\begin{figure}
\centering
\begin{tabular}{cc}
\includegraphics[height=1.4cm]{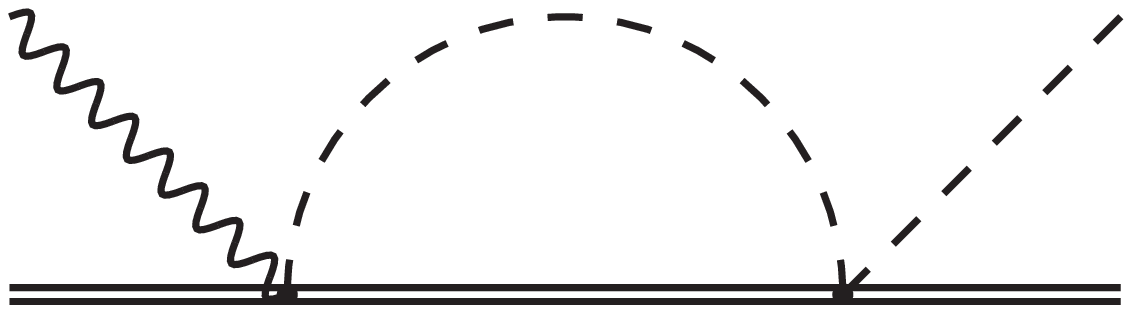}&\includegraphics[height=1.4cm]{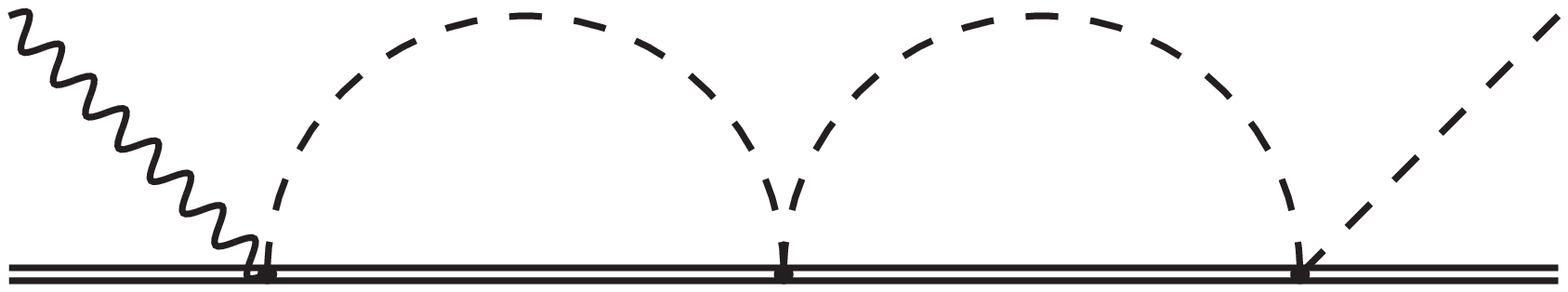}
\end{tabular}
\caption{One- and two loop topologies needed to calculate the
  amplitude. The double line
  generically denotes a nucleon, the dashed line a pion and the
  wiggly line indicates the external electromagnetic field.}\label{fig:diags}
\end{figure}
The one-loop result for channel $(c)$ up to and including order $\O(a \epsilon^4)$ reads
\begin{align} \label{eq:1loop}
\left( \begin{array}{c} E_{0+}^{\mathrm{1 Loop}}(s) \\ \frac{1}{|\vq|}M_{1+}^{\mathrm{1
      Loop}}(s) \\ \frac{1}{|\vq|} M_{1-}^{\mathrm{1 Loop}}(s)
  \\ \frac{1}{|\vq|} E_{1+}^{\mathrm{1 Loop}}(s) \end{array}
  \right) &= \left( \begin{array}{cc}  P^{(c)}_{11} & P^{(c)}_{12}\\ P_{21}^{(c)}  & P^{(c)}_{22} \\ P^{(c)}_{31} &
     P^{(c)}_{32} \\ P^{(c)}_{41} &
     P^{(c)}_{42}  \end{array}
  \right) \left( \begin{array}{c}  J_{ab}(s) \\ J_{cd}(s)
     \end{array} \right) \co
\end{align}
where $m_a,M_{\pi^b}$ denote the masses of the final state of the pertinent
channel and $m_c,M_{\pi^d}$ stands for the masses of the intermediate
state that differ from the final state masses.
The elements $P^{(c)}_{ik}$ are functions of the pion momentum $\vq$
and the coupling constants of the
Lagrangian. For the channel $(p0)$, one finds
\begin{align}\label{eq:1loopcoef}
P_{11}^{(p0)} &= G_0 C_0 + \vq^2 \left(C_0 E^{(p0),t}_{0+,2}-2 D_0^{(2)}
  G_0\right) \co \nn  P_{12}^{(p0)} &= C_1 H_0 + \qstar{2} \left(C_1 E^{(n+),t}_{0+,2}-D_1^{(2)} H_0 \right)-\vq^2 D_1^{(2)} H_0 \co\nn
18 P_{21}^{(p0)} &= \vq^2 \left(D^{(1)}_0-D^{(3)}_0\right)( G_1-G_3) \co\nn
18 P_{22}^{(p0)} &=  \qstar{2} \left(D^{(1)}_1-D^{(3)}_1 \right) (
H_1-H_3) \co \nonumber
\end{align}
\begin{align}
9 P_{31}^{(p0)} &= \vq^2 \left(D^{(1)}_0+2D^{(3)}_0 \right)(G_3-G_1-3G_2) \co\nn
9 P_{32}^{(p0)} &= \qstar{2} \left(D^{(1)}_1+2D^{(3)}_1\right)(H_3-H_1-3H_2)\co \nn
18 P_{41}^{(p0)} &= \vq^2 \left(D^{(1)}_0-D^{(3)}_0\right)(G_1+G_3) \co\nn
18 P_{42}^{(p0)} &= \qstar{2} \left(D^{(1)}_1-D^{(3)}_1\right)(H_1+H_3) \co
\end{align}
where $E^{(c),t}_{0+,2}$ denotes the pertinent coefficient of the tree
level result of channel $(c)$, see Eq.~(\ref{eq:tree}), and $\qstar{2}$ is given by
\begin{align}
\qstar{2} &= \frac{\left(s-(m_c+M_{\pi^d})^2\right)\left(s-(m_c-M_{\pi^d})^2\right)}{4s} \co
\end{align}
which is a quantity of order $\epsilon^2$. Eq.~(\ref{eq:1loop}) and
(\ref{eq:1loopcoef}) clearly show the advantage of the nonrelativistic
description: The strength of the cusp in the channel $(p0)$ is parameterized in terms of the
coupling constant $C_1$ and the ratio $H_0/G_0$.

{\bf 9.} The two-loop corrections all have the topology shown in
Fig.~\ref{fig:diags} and can therefore be cast into the form
\begin{align} \label{eq:2loop}
E_{0+}^{\mathrm{2 Loop}}(s) &= \big( J_{ab}(s) \,\,\, J_{cd}(s) \big) \left( \begin{array}{cc}  T^{(c)}_{11} & T^{(c)}_{12}\\ T_{12}^{(c)}  & T^{(c)}_{22}  \end{array}\right) \left( \begin{array}{c}  J_{ab}(s) \\ J_{cd}(s)
     \end{array} \right) \co
\end{align}
where the $T^{(c)}_{ij}$ for the channel $(p0)$ read
\begin{align}
T_{11}^{(p0)} &= C^2_0 G_0 +C^2_0 E^{(p0),t}_{0+,2}  \vq^2 -4 C_0 G_0
D_0^{(2)} \vq^2 \co\nn
T_{12}^{(p0)} &= \frac{1}{2}C_1^2\,G_0 + 
  \frac{1}{2}C_0\,C_1\,H_0 + \frac{1}{2}C_1^2\, E^{(p0),t}_{0+,2} \vq^2
    - C_1\,H_0\,
   D_0^{(2)}\vq^2 - 
  C_1\,G_0\,D_1^{(2)} \vq^2 \nn &-
      \frac{1}{2} C_0\,H_0\,
     D_1^{(2)}\vq^2 + 
  \frac{1}{2} C_0\,C_1\,E^{(n+),t}_{0+,2}\,
     \qstar{2}\nn &- 
  C_1\,G_0\,D_1^{(2)} \, \qstar{2}
  - \frac{1}{2} C_0\,H_0\,
     D_1^{(2)}\, \qstar{2} \co\nn
T_{22}^{(p0)} &= C_1 C_2 H_0 - C_2 H_0   D_1^{(2)}\vq^2 +
    C_1 C_2  E^{(n+),t}_{0+,2} \qstar{2}\nn  &-
    C_2  H_0 D_1^{(2)} \qstar{2} -
    2 C_1 H_0 D_2^{(2)} \qstar{2} \fs
\end{align}
Up to and including order $\O(a^2 \epsilon^5)$, the two-loop
corrections are independent of the scattering angle $\cos\Theta$ and
only contribute to the amplitude $\Fnr_1$. Therefore, at the order
considered here, the electric multipole $E_{0+}$ is the only quantity
which receives two-loop corrections.

{\bf 10.} The result for the other channels are obtained from the tree
level result of channel $(p0)$ and the coefficients $P_{ij}^{(p0)}$ and $T_{ij}^{(p0)}$
by the replacements
\begin{align}
(n+) &: \{G_i,H_i,C_0,C_2 \} \to
  \{H_i,G_i,C_2,C_0 \}\co \nn
(n0) &: \{G_i,H_i,C_0,C_1,C_2 \}   \to
  \{L_i,K_i,C_3,C_4,C_5 \} \co \nn
(p-) &: \{G_i,H_i,C_0,,C_1,C_2 \}   \to
  \{K_i,L_i,C_5,C_4,C_3 \} \fs
\end{align}
The replacement indicated for the $C_x$ has to be done also for the
corresponding $D^{(i)}_x$. 

{\bf 11.} In an isospin symmetric world, the phase of the multipole $E_{0+}$ is
directly related to the phase shift of the $S$-wave of pion-nucleon scattering
by virtue of the Fermi-Watson theorem \cite{FW}. In
Ref.~\cite{bernstein}, it is shown with a coupled channel $S$-matrix
approach that in the channel $(p0)$, to leading order in $e$, below the $\pi^+ n$
 threshold, the phase of the $S$ wave of $\pi^0 p \to \pi^0 p$
 scattering is equal to the phase of $E_{0+}$,
\begin{align}
\tan \delta_{p \pi^0 \to p \pi^0} &= \tan
\frac{\mathrm{Im\,}E_{0+}}{\mathrm{Re\,}E_{0+}} \equiv \tan \delta_{E_{0+}}  \fs
\end{align}
The framework developed here allows one to test this statement order
by order in the perturbative expansion. To this end, the phase of $E_{0+}$ below the second
threshold is calculated up to and including $\O(a^2 \eps^4)$,
\begin{align}\label{eq:phase}
\tan\delta_{E_{0+}} &= C_0\, \mr{Im}J_{p0} + C_1^2 J_{n+} \mr{Im} J_{p0} - 2D_0^{(2)}
\mr{Im}J_{p0} \vq^2 \\& -2C_1 D_1^{(2)}J_{n+} \mr{Im} J_{p0} \, \vq^2 - 2C_1 D_1^{(2)}
J_{n+} \mr{Im}J_{p0} \, h^2(s,\mn,\mpc) + \cdots \fs \nonumber
\end{align}
Calculating $\pi^0 p \to \pi^0 p$ scattering to the same order with
the Lagrangian given in Eq.~(\ref{eq:Lpn}), one
finds that the phase of the $S$-wave below the second threshold is indeed equal to
Eq.~(\ref{eq:phase}). However, the main object of interest is the
phase of the $S$-wave of $\pi^0 p \to \pi^0 p$
scattering in the {\it isospin symmetry limit},
\begin{align}
\tan \bar{\delta}_{p\pi^0 \to p\pi^0} &= C_0 \mr{Im}J_{p0} + \frac{C_1^2}{C_0}
\mr{Im}J_{n+} \simeq a_{0+}^+ \vq +2\frac{a_{0+}^{-2}}{a_{0+}^+}\vq + \cdots \co
\end{align}
which does not agree with
$\delta_{p\pi^0\to p\pi^0}$ in the presence of isospin violations
already at leading order.

{\bf 12.} In this letter, the photoproduction reaction of pions on the 
  nucleon is studied using a nonrelativistic framework. The electric
  and magnetic multipoles $E_{l+}$ for $l = 0,1$ and $M_{1\pm}$ are
  calculated in a systematic double expansion in the final state pion-
  and nucleon momenta (counted as a small quantity of order $\epsilon$) and
  the threshold parameters of $\pi N$ scattering (denoted by
  $a$). Explicit representations for the multipoles up to and
  including $\epsilon^3$, $\epsilon^4 a$, $\epsilon^4 a^2$ are
  provided.  The representation is valid in the low energy region,
  at least up to a photon energy in the lab frame of $E_\gamma = 165
  \MeV$. It accurately describes the cusp structure and allows one to
  determine the pion-nucleon threshold parameters from experimental data.

  The relation of the phase of the electric multipole $E_{0+}$ in the
  $(p0)$ channel to the phase of the $S$-wave of $\pi^0 p \to \pi^0 p$
  scattering is discussed in the presence of isospin violation. A
  relation found in earlier work \cite{bernstein} is
  confirmed. We stress that the relation does not allow one to obtain
  the phase of $\pi^0 p \to \pi^0 p$ scattering in the isospin limit.

{\it Acknowledgements.} I would like to thank J.~Gasser, B.~Kubis, A.~Manohar and
U.-G.~Mei\ss{}ner for informative discussions and B.~Kubis for
comments on the manuscript. This work was supported
in part by the Department
of Energy under Grant 
DE-FG03-97ER40546 and by the Swiss National Science Foundation.

\renewcommand{\thefigure}{\thesection.\arabic{figure}}
\renewcommand{\thetable}{\thesection.\arabic{table}}
\renewcommand{\theequation}{\thesection.\arabic{equation}}

\appendix

\setcounter{equation}{0}
\setcounter{figure}{0}
\setcounter{table}{0}

\section{Matching relations}\label{app:matching}

The matching relations of the nonrelativistic couplings to the
threshold parameters defined in Eq.~(\ref{eq:thresholdpar}) read
\begin{align}
G_4 &= 15 (\bar{E}_{2+}+2 \bar{M}_{2+}) \co \nn
G_5 &= 2\zeta \bar{E}_{0+}- 2\bar{E}_{2-}- \kappa \bar{E}_{0+} -2
\bar{E}_{0+,2}+3 (\bar{E}_{2+}-2
  \bar{M}_{2-}+2\bar{M}_{2+}) \co \nn
G_6 &= -6 (3\bar{M}_{2+}+2\bar{M}_{2-}) \co \nn
G_7 &= 30 (\bar{E}_{2+}-\bar{M}_{2+}) \co  \nn
G_8 &= 6 (\bar{E}_{2-} -
  \bar{M}_{2+}+\bar{M}_{2-}+\bar{E}_{2+})\co \nn
 G_9 &= 2G_1 \zeta - 6\bar{E}_{3-}-3 \kappa
 \bar{E}_{1+}-6\bar{E}_{1+,2}+16\bar{E}_{3+}-24\bar{M}_{3-}-3 \kappa
 \bar{M}_{1+}\nn
&-6 \bar{M}_{1+,2}+ 45 \bar{M}_{3+}\co \nn
 G_{10} &= 35 (\bar{E}_{3+}+3\bar{M}_{3+})\co \nn
G_{11} &= G_2 \zeta + \kappa \bar{M}_{1-}+ 2\bar{M}_{1-,2}-
9\bar{M}_{3-}+2\kappa \bar{M}_{1+} + 
  4 \bar{M}_{1+,2} -12\bar{M}_{3+} \co \nn
G_{12} &= -15 (3\bar{M}_{3-}+4\bar{M}_{3+}) \co  \nn
G_{13} &= 2 G_3 \zeta -6 \bar{E}_{3-}-3 \kappa \bar{E}_{1+}-6
\bar{E}_{1+,2}+15\bar{E}_{3+}-6 \bar{M}_{3-}+ 3 \kappa \bar{M}_{1+}
\nn &+ 6 \bar{M}_{1+,2}- 15 \bar{M}_{3+}\co \nn
G_{14} &= 105 (\bar{E}_{3+}-\bar{M}_{3+}) \co  \nn
G_{15} &= 30
  (\bar{M}_{3-}+\bar{E}_{3-}+\bar{E}_{3+}-\bar{M}_{3+})\co
\end{align}
with $\kappa = \frac{1}{\mpn \mpr}$ and $\zeta = \frac{k_1}{k_0}$.

\end{document}